\documentclass[]{spie}  
\usepackage[]{graphicx}

\def\aj{AJ}%
\def\apj{ApJ}%
\def\apjs{ApJS}%
\def\ao{Appl.~Opt.}%
\def\aap{A\&A}%
\def\mnras{MNRAS}%
\def\pasp{PASP}%

\title{Exoplanet science with the LBTI: instrument status and plans}

\author{D.~Defr\`ere\supit{a}, P.~Hinz\supit{a}, A.~Skemer\supit{a}, V.~Bailey\supit{a}, E.~Downey\supit{a}, O.~Durney\supit{a}, J.~Eisner\supit{a}, J.M.~Hill\supit{b}, W.F.~Hoffmann\supit{a}, J.~Leisenring\supit{a}, T.~McMahon\supit{a}, M.~Montoya\supit{a}, E.~Spalding\supit{a}, J.~Stone\supit{a}, A.~Vaz\supit{a}, O.~Absil\supit{c}, S.~Esposito\supit{d}, M.~Kenworthy\supit{e}, B.~Mennesson\supit{f}, R.~Millan-Gabet\supit{g},  M.~Nelson\supit{h}, A.~Puglisi\supit{d}, M.F.~Skrutskie\supit{i}, J.~Wilson\supit{i}
\skiplinehalf
\supit{a}Steward Observatory, University of Arizona, 933 N. Cherry Avenue, 85721 Tucson, USA\\
\supit{b}Large Binocular Telescope Observatory, University of Arizona, 933 N. Cherry Avenue, 85721 Tucson, USA\\
\supit{c}D\'epartement d'Astrophysique, G\'eophysique et Oc\'eanographie, Universit\'e de Li\`ege, 17 All\'ee du Six Ao\^ut, B-4000 Sart Tilman, Belgium\\
\supit{d}Arcetri Astrophysical Observatory, Largo Enrico Fermi 5, 50125 Firenze, Italia\\
\supit{e}Leiden Observatory, Niels Bohrweg 2, 2300 RA, Leiden, The Netherlands\\
\supit{f}Jet Propulsion Laboratory, California Institute of Technology 4800 Oak Grove Drive, Pasadena CA 91109-8099, USA\\
\supit{g}NASA Exoplanet Science Center (NExSci), California Institute of Technology, 770 South Wilson Avenue, Pasadena CA 91125, USA\\
\supit{h}University of Minnesota, 116 Church St SE, Minneapolis, NM, 55455, USA\\
\supit{i}University of Virginia, Department of Astronomy, 530 McCormick Road, Charlottesville, VA 22904-4325, USA\\
}

\authorinfo{Send email correspondence to D.D. at ddefrere@email.arizona.edu.}

\begin{document} 
\maketitle 

\begin{abstract}
The Large Binocular Telescope Interferometer (LBTI) is a strategic instrument of the LBT designed for high-sensitivity, high-contrast, and high-resolution infrared (1.5-13~$\mu$m) imaging of nearby planetary systems. To carry out a wide range of high-spatial resolution observations, it can combine the two AO-corrected 8.4-m apertures of the LBT in various ways including direct (non-interferometric) imaging, coronagraphy (APP and AGPM), Fizeau imaging, non-redundant aperture masking, and nulling interferometry. It also has broadband, narrowband, and spectrally dispersed capabilities. In this paper, we review the performance of these modes in terms of exoplanet science capabilities and describe recent instrumental milestones such as first-light Fizeau images (with the angular resolution of an equivalent 22.8-m telescope) and deep interferometric nulling observations. 
\end{abstract}

\section{INTRODUCTION}
\label{sec:intro}  

The Large Binocular Telescope Interferometer\cite{Hinz:2014} (LBTI) is a NASA-funded nulling and imaging instrument that combines the two primary mirrors of the Large Binocular Telescope\cite{Hill:2014,Veillet:2014} (LBT) for high-sensitivity, high-contrast, and high-resolution infrared imaging (1.5-13\,$\mu$m). It is designed to be a versatile instrument that can image the two aperture beams separately (see Section~\ref{sec:leech}), overlap the two beams incoherently (see Section~\ref{sec:leech}), overlap the beams coherently (in phase) for Fizeau interferometry (see Section~\ref{sec:fizeau}), or overlap the beams and pupils for nulling interferometry (see Section~\ref{sec:hosts}). Various other specialized modes relevant for exoplanet science are also available such as coronagraphy (see Section~\ref{sec:agpm}), non-redundant aperture masking (see Section~\ref{sec:nrm}), and an integral field spectrograph (see Section~\ref{sec:ales}). While these modes are currently mostly used in single-aperture mode (i.e., with the two beams separated on the detector), they can also take advantage of beam combination to maximize sensitivity (incoherent combination) or angular resolution (coherent combination). In coherent mode, the LBTI provides a spatial resolution equivalent to that of a 22.8-meter telescope along the horizontal/azimuthal axis and the light-gathering power of single 11.8-meter mirror which position the LBT as a forerunner of the new generation of extremely large telescopes (ELT). This makes the LBTI a very unique instrument that can tackle a wide range of observing programs and challenges. In this paper, we describe several technical milestones achieved over the past few years and show some key science results with a particular focus on the two main ongoing survey: the planet survey called LEECH (LBTI Exozodi Exoplanet Common Hunt, see Section~\ref{sec:leech}) and the exozodiacal dust survey called HOSTS (Hunt for Observable Signatures of Terrestrial Planetary Systems, see Section~\ref{sec:hosts}).

\begin{figure}[!t]
	\begin{center}
		\includegraphics[width=14.5 cm]{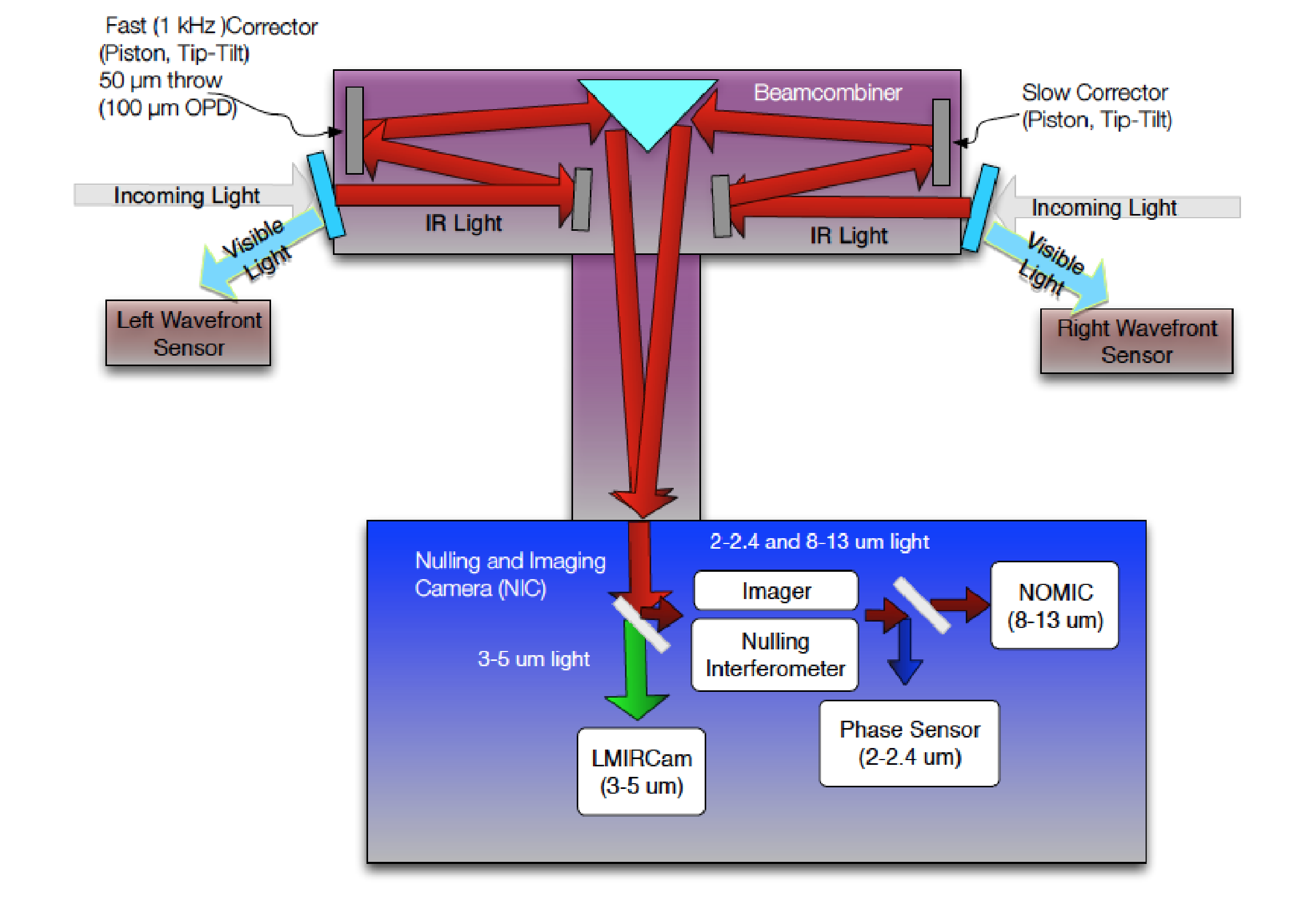}
		\caption{Components of the LBTI shown with the optical path through the beam combiner and the NIC cryostat. Starlight is reflected on LBT primaries, secondaries, and tertiaries before coming into this diagram on the top right and top left. The visible light is reflected on the entrance window and used for adaptive optics while the infrared light is transmitted into LBTI, where all subsequent optics are cryogenic. The beam combiner directs the light with steerable mirrors and can adjust pathlength for interferometry. Inside the NIC cryostat, 3-5$\mu$m light is directed to LMIRCam for exoplanet imaging, 2.0-2.4$\mu$m light is directed to the phase sensor, which measures the differential tip/tilt and phase between the two primary mirrors, and 8-13$\mu$m light is directed to NOMIC for Fizeau imaging or nulling interferometry.}\label{fig:diagram}
	\end{center}
\end{figure}
  
\section{INSTRUMENT}
\label{sec:instr}  

The LBTI is located at the bent center Gregorian focal station of the LBT. The LBT is a two 8.4-m aperture optical telescope installed on Mount Graham in southeastern Arizona (at an elevation of 3192 meters) and operated by an international collaboration between institutions in the United States, Italy, and Germany. Both apertures are equipped with state-of-the-art high-performance adaptive optics systems to correct for atmospheric turbulence\cite{Esposito:2010,Bailey:2014}. Each deformable secondary mirror uses 672 actuators that routinely correct 500 modes and provide Strehl ratios of 80\%, 95\%, and 99\% at 1.6\,$\mu$m, 3.8\,$\mu$m, and 10\,$\mu$m, respectively\cite{Esposito:2012,Skemer:2014}. After bouncing off of the LBT primaries, secondaries, and tertiaries, starlight enters the LBTI as represented in Figure~\ref{fig:diagram}. Visible light reflects first off on the LBTI entrance windows and into the adaptive optics wavefront sensors, which control the deformable secondary mirror. Infrared light transmits into LBTI's universal beam combiner (UBC) from which all optics are cryogenic. The UBC directs the light with steerable mirrors to the cryogenic Nulling Infrared Camera (NIC) which is equipped with two scientific cameras: 

\begin{table}[!t]
\caption{List of LBTI's filters currently available for LMIRCam and NOMIC observations.}
\begin{center}
\begin{tabular}{c c c |c c c}
\hline
\hline
        &  LMIRCam        &           &         & NOMIC           &            \\
Name    & $\lambda_{\rm eff}$ [$\mu$m] & FWHM [$\mu$m] & Name    & $\lambda_{\rm eff}$ [$\mu$m] & FWHM [$\mu$m]\\
\hline
FeII    &  1.645          &  0.03     & N07904-9N  &  7.904     & 0.700      \\
H       &  1.655          &  0.31     & W08699-9   &  8.699     & 1.120      \\
H2-on   &  2.125          &  0.03     & N08909-9O  &  8.909     & 0.760      \\   
K       &  2.160          &  0.32     & N09145-9   &  9.145     & 0.810      \\
H2-off  &  2.255          &  0.03     & N09788-9P  &  9.788     & 0.920      \\
L-NB1   &  3.040          &  0.15     & W10288-8   &  10.29     & 6.010      \\
L-NB2   &  3.160          &  0.08     & W10550-9Q  &  10.55     & 0.970      \\
PAH1    &  3.285          &  0.05     & N11855-8R  &  11.85     & 1.130      \\
L-NB3   &  3.310          &  0.16     & N'         &  11.10     & 2.600      \\   
$[$3.3$\mu$m$]$ &  3.310      &  0.40     & N12520-9S  &  12.52     & 3.160      \\   
Lspec   &  3.405          &  1.19     &    &            &            \\
L-NB4   &  3.465          &  0.15     &    &            &            \\
L-NB5   &  3.595          &  0.09     &    &            &            \\
L       &  3.700          &  0.58     &    &            &            \\
L-NB6   &  3.705          &  0.19     &    &            &            \\
L-cont4 &  3.780          &  0.20     &    &            &            \\
L-NB7   &  3.875          &  0.23     &    &            &            \\
L-NB8   &  4.000          &  0.06     &    &            &            \\
Br-$\alpha$-off & 4.005   &  0.07     &    &            &            \\
Br-$\alpha$-on & 4.055    &  0.07     &    &            &            \\
M       &  4.785          &  0.37     &    &            &            \\
\hline
\end{tabular}
\end{center}
\label{tab:filters}
\end{table}

\begin{itemize}
\item LMIRCam\cite{Skrutskie:2010,Leisenring:2012} (the L and M Infrared Camera) is the mid-infrared optimized science camera equipped with a Hawaii-2RG HgCdTe detector which has a very fine plate-scale (0.0107 arcsec/pixel) that Nyquist samples a K-band interferometric PSF and over-samples single aperture adaptive optics images. The current electronics read a sub-array of 1024x1024 pixels but will be upgraded soon to read the full detector (2048x2048 pixels). LMIRCam contains various filter options spanning 1-5$\mu$m, including for instance the L band, M band, PAH-on, PAH-off, Br$\alpha$, and H$_2$O ice (see list of available filters in table~\ref{tab:filters}). In addition, LMIRCam has a set of germanium grisms\cite{Kuzmenko:2012} (R$\sim$400), two vector-vortex coronagraphs (see Section~\ref{sec:agpm}), a set of Apodizing Phase Plate (APP) coronagraphs\cite{Kenworthy:2010}, and an integral field spectrograph, all of which are in the testing phase and not currently used for the LEECH survey.

\item NOMIC\cite{Hoffmann:2014} (Nulling Optimized Mid-Infrared Camera) is the long-wavelength camera used in particular for nulling interferometry. The detector is a 1024x1024 Blocked Impurity Band (BIB) hybrid array with 30-$\mu$m pixels. The optics provides a field of view of 12~arcsecs with pixels of 0.018~arcsecs in size. $\lambda$/D for an individual aperture is 0.27~arcsecs or 0.10 arcsecs (or 5.5 pixels) for Fizeau interferometry (at 11$\mu$m).

\end{itemize}

Finally, NIC also contains a near-infrared camera equipped with a PICNIC detector for relative tip/tilt and fringe sensing (PHASECam). PHASECam operates at 1kHz and receives the light from both interferometric outputs whether the long wavelength channel is in nulling or imaging mode. Beam alignment is done via the Fast Pathlength Corrector (FPC), located in the left part of the UBC, and the Slow Pathlength Corrector (SPC), located in the right side. Both the FPC and the SPC can adjust pathlength for interferometry. The FPC provides a Piezo-electric transducer (PZT) fast pathlength correction with a 80\,$\mu$m of physical stroke, capable of introducing 160\,$\mu$m of optical path difference (OPD) correction. The right mirror provides a larger stroke (40 mm of motion) for slow pathlength correction. In practice, the SPC is used to acquire the fringes while the FPC is used to correct for pathlength variations at high speed. The LBT is an ideal platform for interferometric observations because both telescopes are installed on a single steerable mount. This design does not require long delay lines and contains relatively few warm optical elements, which provides an exceptional sensitivity. More information about the design and performance of PHASECam can be found in Defr\`ere et al. (2014)\cite{Defrere:2014c}.

\begin{figure}[!b]
	\begin{center}
		\includegraphics[width=8.5cm]{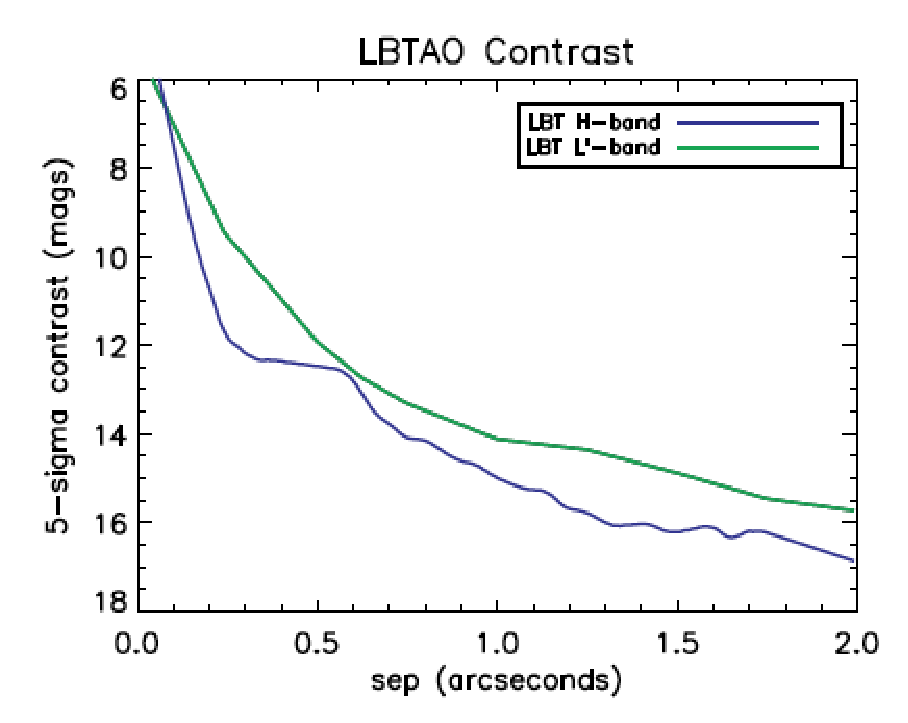}
		\includegraphics[width=8.5cm]{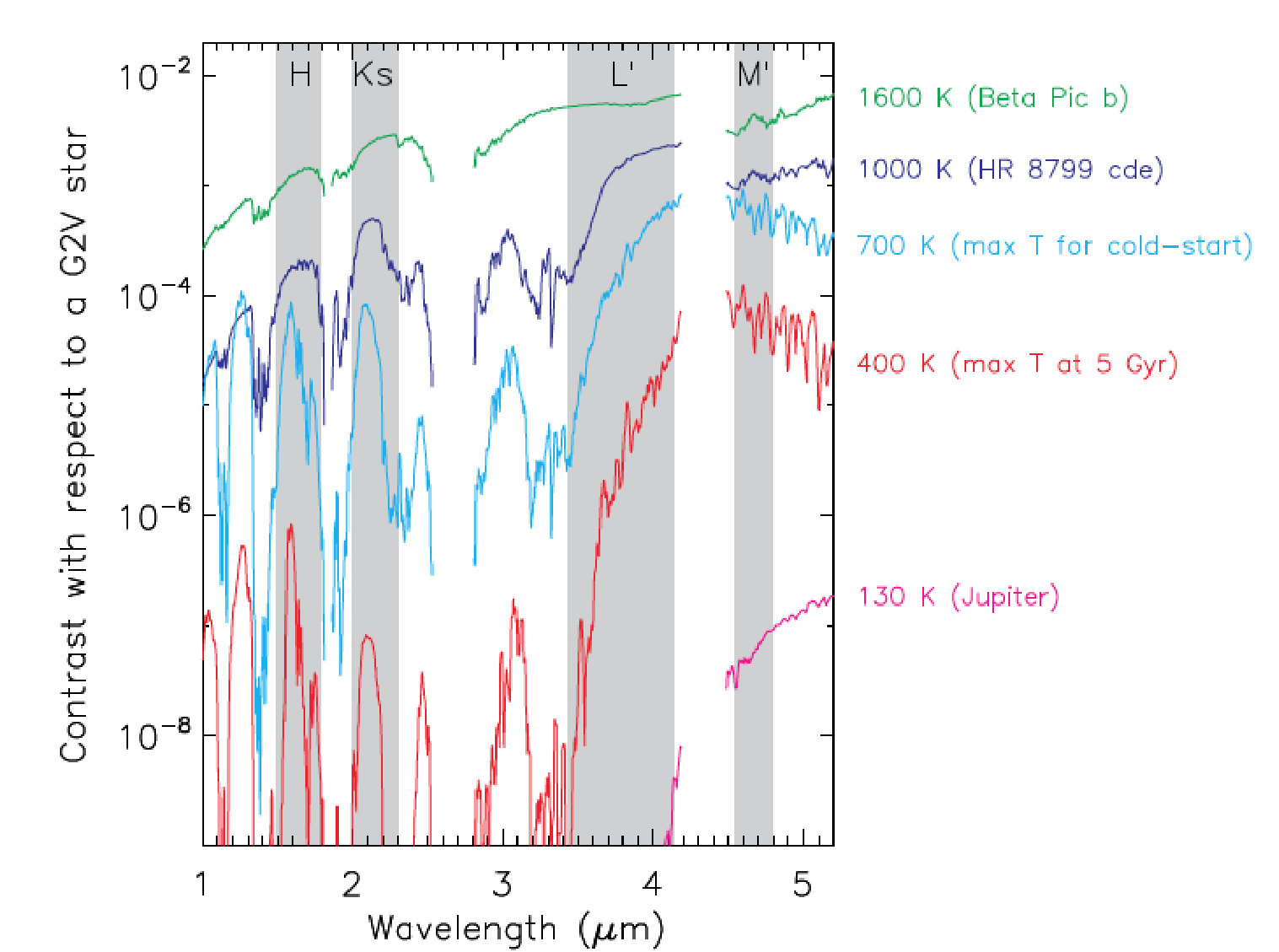}
		\caption{Left, on-sky contrast curves for LBTAO at H and LÕ band (1.6\,$\mu$m and 3.8\,$\mu$m respectively). Right, characteristic examples of exoplanet-to-star contrasts (i.e. flux ratios) as a function of wavelength, showing (1) that gas-giant exoplanets can be detected with lower contrasts in the mid-infrared (3-5\,$\mu$m) than in the near-infrared (1-2\,$\mu$m), and (2) that this difference increases at lower temperatures (Figure from Skemer et al.\ 2014\cite{Skemer:2014}).}\label{fig:leech}
	\end{center}
\end{figure}

\section{The LEECH planet survey}\label{sec:leech}

The LBTI Exozodi Exoplanet Common Hunt\cite{Skemer:2014} (LEECH) is a large multinational survey ($\sim$100 nights) to search for and characterize directly imaged exoplanets in the mid-infrared (3-5\,$\mu$m). LEECH offers key and unique advantages for exoplanet detection and characterization. First, giant exoplanet fluxes peak between 4 and 5\,$\mu$m (see Figure~\ref{fig:leech}, right) so that lower-mass and older objects can be observed compared to those observed shorter wavelengths. Second, AO systems perform better at longer wavelengths, which reduces speckle noise and improves the sensitivity at small angular separation. For these reasons, LEECH's contrast is competitive with, and neatly complements, other high-contrast planet imaging efforts focused on near-infrared bands (1-2.4\,$\mu$m).

The LEECH survey began its 100-night campaign in Spring 2013 and is currently mainly focused on exoplanet detection at L' (3.8\,$\mu$m) using the LBTI in single-aperture mode (taking advantage of the outstanding performance of LBT's adaptive optics system). The sensitivity and utility of the LBTAO coupled with LBTI/LMIRCam has been demonstrated in several studies of known, substellar companions and a very low-mass binary\cite{Skemer:2012, Bonnefoy:2014, Schlieder:2014}. These capabilities also led to strong constraints on the possibility of fifth planet in the HR\,8799 planetary system during the LEECH survey\cite{Maire:2015}. The typical on-sky contrast curve obtained by LEECH is represented in Figure~\ref{fig:leech} (left) and shows the exquisite performance achieved by the instrument. Also shown in this Figure is the H-band contrast which is superior to the L' band one but becomes inferior when converted to mass contrasts using models\cite{Baraffe:1998}. More information about the survey design and the target list can be found in Skemer et al.\ (2014)\cite{Skemer:2014}.

LEECH is also used to characterize known directly imaged exoplanets from 3 to 5\,$\mu$m\cite{Skemer:2012,Skemer:2014b}, a wavelength range that contains the methane fundamental absorption feature and where previous work has struggled to explain planet SEDs with atmospheric models. For that purpose, LEECH can take advantage of LBTI's versatility by using a combination of different filters and incoherently overlap the two-aperture beams as shown in recent studies of the HR\,8799bcde\cite{Skemer:2014b} and GJ\,504b (Skemer et al.\ in prep) systems. While other exciting results are in the pipeline, a major step toward mid-infrared exoplanet characterization was achieved recently with the first light of LBTI's new integral field spectrograph (IFS, see Section~\ref{sec:ales}). It is the first IFS operating in the mid-infrared (from 3 to 5\,$\mu$m) and will considerably extend our ability to characterize self-luminous exoplanets at wavelengths critical for planning observations and advancing our theoretical understanding of self-luminous planets in advance of JWST.

\begin{figure}[!b]
	\begin{center}
		\includegraphics[width=8.3 cm]{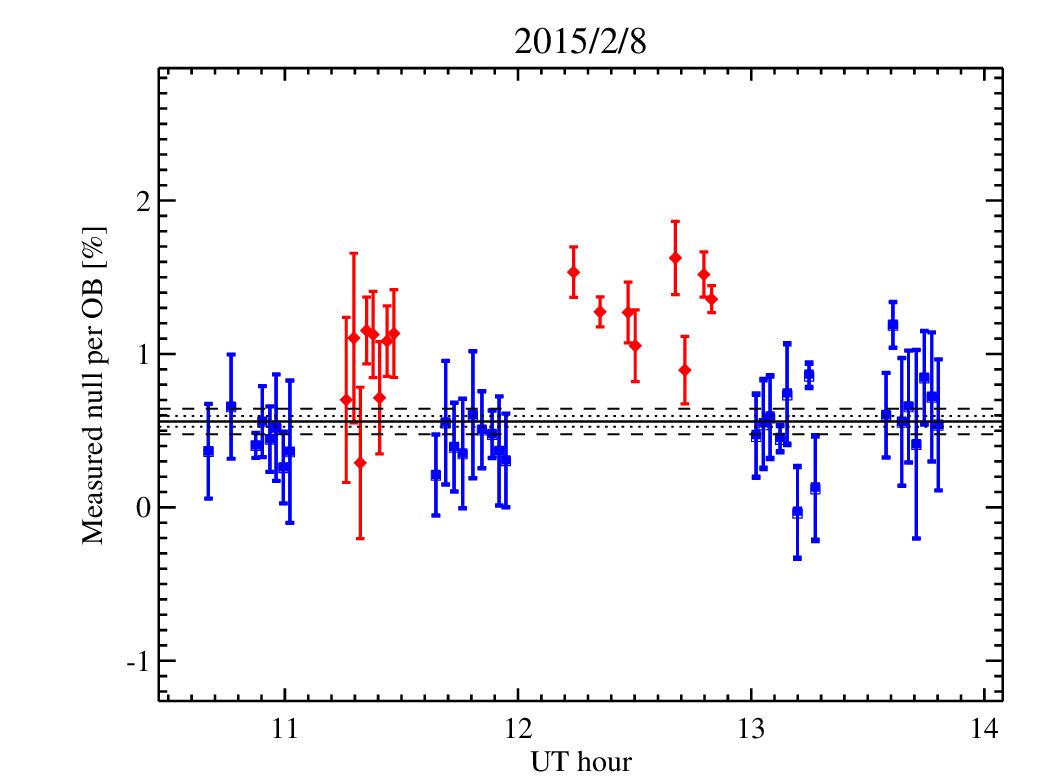}
		\includegraphics[width=8.7 cm]{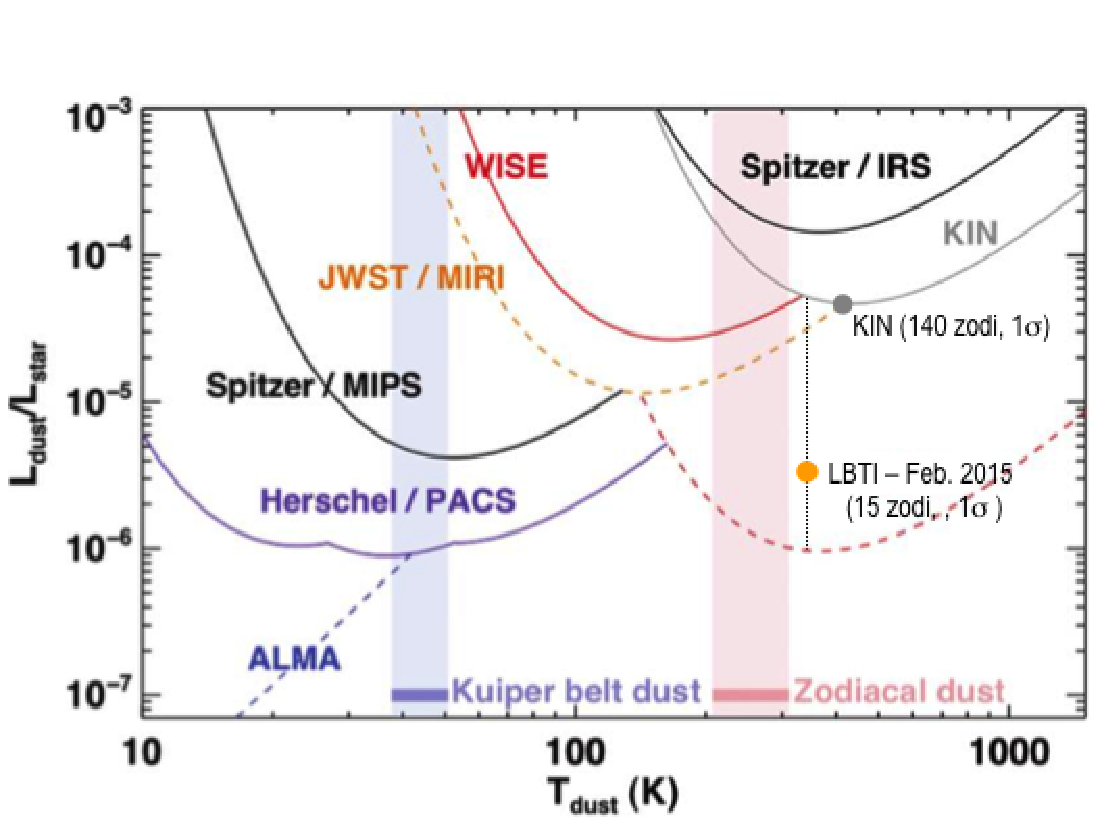}
		\caption{Left, raw null measurements per observing block as a function of UT time obtained on February 8, 2015 and showing a calibrated null accuracy of 0.05\%. The blue squares show the calibrator measurements while the red diamonds represent the $\beta$\,Leo measurements. The estimated instrumental null floor is represented by the solid black line and the corresponding 1-$\sigma$ uncertainty by the dashed lines. Right, corresponding LBTI sensitivity limit for detection of debris dust around nearby Sun-like stars compared to various recent (Spitzer, WISE, KIN), current (Herschel, ALMA) and near-term facilities (JWST). The curves show 3-$\sigma$ detection limits in terms of the fractional dust luminosity (L$_{\rm dust}$=L$_\star$) vs. its temperature (Figure adapted from Roberge et al.\ 2012\cite{Roberge:2012}).}\label{fig:null}
	\end{center}
\end{figure}

\section{The HOSTS exozodi survey}\label{sec:hosts}

The Hunt for Observable Signatures of Terrestrial planetary Systems\cite{Hinz:2013,Danchi:2014} (HOSTS) program on the LBTI will survey nearby stars for faint exozodiacal dust (exozodi). This warm circumstellar dust, analogous to the interplanetary dust found in the vicinity of the Earth in our own system, is produced in comet breakups and asteroid collisions. Emission and/or scattered light from the exozodi will be the major source of astrophysical noise for a future space telescope aimed at direct imaging and spectroscopy of terrestrial planets (exo-Earths) around nearby stars\cite{Defrere:2010,Roberge:2012}. The prevalence of exozodiacal dust in the terrestrial planet region of nearby planetary systems is currently poorly constrained and must be determined to design these future space-based instruments. So far, only the bright end of the exozodi luminosity function has been measured on a statistically meaningful sample of stars\cite{Kennedy:2013}. To determine the prevalence of exozodiacal dust at the faint end of the luminosity function, NASA has funded the Keck Interferometer Nuller (KIN) and the Large Binocular Telescope Interferometer (LBTI) to carry out surveys of nearby main sequence stars. Science results from the KIN were reported recently\cite{Millan-gabet:2011,Mennesson:2014} and show a typical sensitivity per object approximately one order of magnitude larger than that required to prepare future exo-Earth imaging instruments. The LBTI is designed to reach the required level. The first scientific result, based on commissioning data, was reported recently and shows the detection of exozodiacal dust around $\eta$~Crv\cite{Defrere:2015}. These observations were made using a very coarse fringe tracking algorithm, i.e. the so-called CG algorithm\cite{Defrere:2014} equivalent to group delay tracking, which is limited to a precision of approximately 1\,$\mu$m RMS. For that reason, the nulling data were reduced by keeping only the 5\% of the frames that show the lowest flux. In February 2015, nulling observations of $\beta$~Leo were obtained using phase tracking which was being commissioned and improves considerably the phase stability of the system ($\sim$350nm RMS). This also allowed us to use a much more powerful nulling data reduction analysis, the so-called Nulling Self Calibration or NSC\cite{Hanot:2011,Mennesson:2011}. The results of these observations are shown in Figure~\ref{fig:null} and demonstrate a null accuracy of 0.05\% which is equivalent to an exozodiacal disk density of 8\,zodis for $\beta$~Leo or 15\,zodis for a Sun-like star located at 10\,pc (using the exozodiacal disk model developed by the HOSTS team\cite{Kennedy:2015}). Following this result, the LBTI successfully went through NASA's operational readiness review in April 2015 and received NASA's green light to enter a one year science validation phase starting this Fall. Overall (including both observations obtained during commissioning and science validation phase), the HOSTS survey will be carried out on a sample of 50 to 60 carefully chosen nearby main-sequence stars\cite{Weinberger:2015}. 

\begin{figure}[!t]
	\begin{center}
		\includegraphics[width=17.0cm]{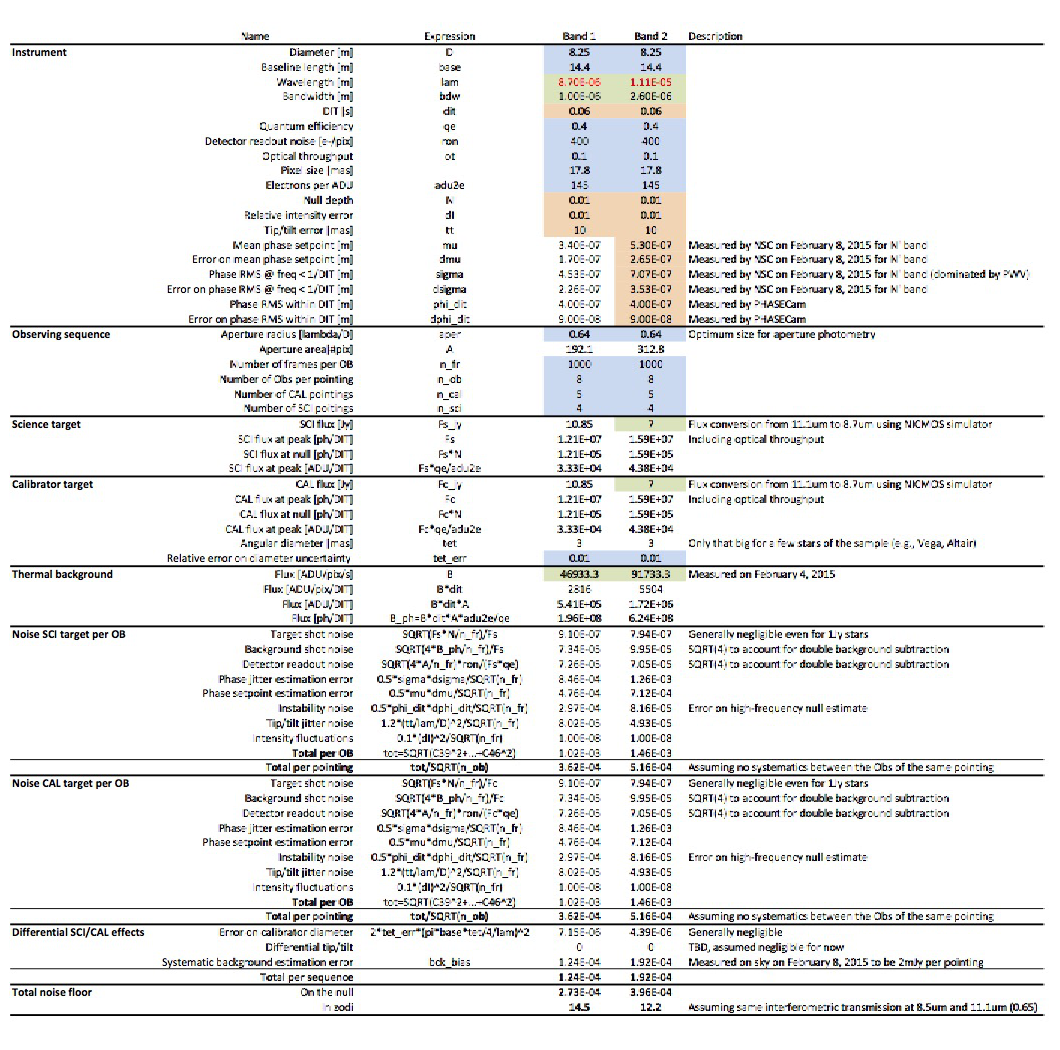}
		\caption{LBTI nulling error budget showing the projected null accuracy for a fully calibrated null sequence ($\sim$3.5 hours) and two wavebands under consideration (i.e., 8.7\,$\mu$m and 11.1\,$\mu$m).}\label{fig:error_budget}
	\end{center}
\end{figure}

One of the top priority during the science validation phase is to address and mitigate the impact of precipitable water vapor, which creates a wavelength-dependent phase between the K-band (where the phase is measured and tracked), and the N' band (where the null measurements are obtained). The KIN encountered the same issue and had to use real-time compensation loops to mitigate this effect. For the LBTI, we are currently working on three different approaches to measure the PWV in real time: (1) using the difference between the group and phase delay at K band, (2) using dispersed fringes obtained with LMIRCam in the mid-infrared, and (3) measuring the two outputs of the fringe tracker in two different near-infrared wavelengths. Such control loops will be implemented and tested during the science validation phase. Also, it is worth noting that the LBTI could also carry out the HOSTS survey at 8.7\,$\mu$m where the impact of PWV is reduced. Unlike the KIN, which got better performance at 8.7\,$\mu$m compared to 11.1\,$\mu$m, the choice is not that obvious for the LBTI because it doesn't have to cope with longitudinal dispersion (because both telescopes are on the same mount) and is not limited by thermal background noise. Another important difference is that 1\,zodi corresponds to a smaller fraction of the stellar flux at 8.7\,$\mu$m (29ppm) compared to 11.1\,$\mu$m (50ppm). Observations at 8.7\,$\mu$m are also more sensitive to phase setpoint and jitter errors. In order to address this issue, we constructed the preceding table (see Figure~\ref{fig:error_budget}) that gives a complete error budget for the two wavebands. Fixed input parameters are marked in blue, inputs depending on integration time in orange, and inputs depending on wavelength in green. The results are given in the last two lines and show that, according to our model, 8.7\,$\mu$m would give a slightly better null accuracy for a full null sequence (270ppm vs 400ppm). However, converting to exozodi luminosity favors 11.1\,$\mu$m (12.2 zodis vs 14.4 zodis). Given the uncertainty in the model (scaling of phase jitter and background bias from 11.1\,$\mu$m to 8.7\,$\mu$m), we cannot give a firm conclusion and obtaining on-sky data with the two wavebands will be one of the top properties of the science validation phase. Other tasks will be focused on reducing the vibrations of the instrument. 

\section{Other results}\label{sec:other}

\subsection{Fizeau imaging}\label{sec:fizeau}

\begin{figure}[!t]
	\begin{center}
		\includegraphics[width=17.0cm]{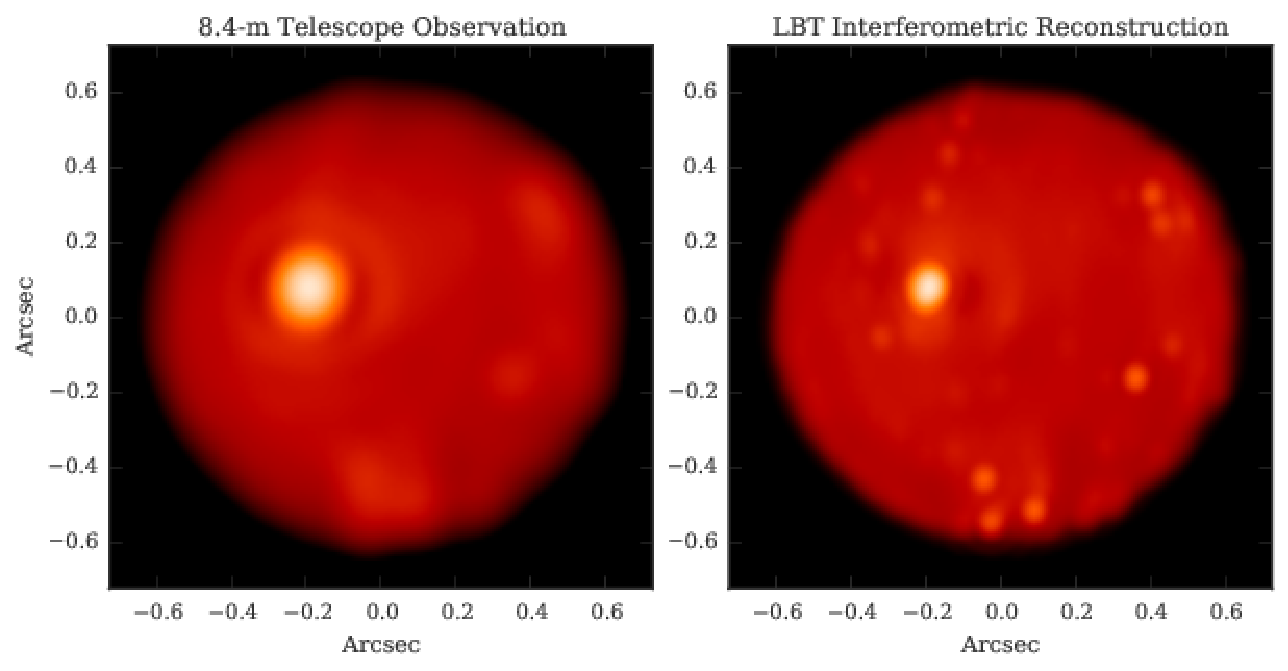}
		\caption{M-band image of Io at the angular resolution equivalent to a single 8.4-meter aperture (left) and interferometric image reconstruction showcasing the angular resolution of the 22.8-m telescope (right).}\label{fig:io}
	\end{center}
\end{figure}

Fizeau interferometric imaging is a powerful technique to get the full angular resolution of the LBT and is available on both LBTI's science channels (separately or simultaneously). The Fizeau PSF can be described by an Airy disk of a single 8.4-meter aperture along the vertical/altitude axis and the product of that Airy disk with a two-slit diffraction pattern along the horizontal/azimuth axis. Since the high angular resolution direction is always parallel to the horizon and perpendicular to the parallactic angle, aperture synthesis is achieved by sky rotation and provides the resolution of 22.8-m circular aperture in post-processing. The capability of this mode has recently been showcased on Jupiter's moon Io (see Figure~\ref{fig:io}). Employing ``lucky Fizeau'' imaging at M-Band, approximately sixteen independent sources, corresponding to known hot spots on the surface of Io and inaccessible to resolutions of 8-meter class telescopes, have been recovered after image reconstruction of the Fizeau interferometric observations\cite{Leisenring:2014,Conrad:2015}.

\subsection{Coronagraphy}\label{sec:agpm}

The LMIRCam channel of the LBTI has a set of coronagraphic devices currently in the testing phase: two vortex coronagraphs\cite{Absil:2014} (AGPM) and two Apodizing Phase Plate coronagraphs\cite{Kenworthy:2010} (APP). First scientific observations with the AGPM vortex were obtained in Fall 2013 during the first hours on sky and revealed the four know planets around HR\,8799 clearly at high SNR\cite{Defrere:2014}. For this first attempt, the data show a very promising starlight rejection ratio of $\sim$35 that can in principle be improved to at least 100 based on our experience with other telescopes. Future work and observations will be focused on improving the rejection ratio using an automatic centering loop based on the science frames and a better algorithm to adjust the focus in the AGPM plane. Note also that a second AGPM was installed in May 2015 and will allow us to observe with both apertures simultaneously, and to get the necessary redundancy which is crucial to distinguish planets from false detections at small angular separations. Combined with LBT's state-of-the-art high-performance adaptive optics system and taking full advantage of LBTI 's extremely low thermal background, we expect LMIRCam/AGPM to open a completely new parameter space for high-contrast imaging at L', a wavelength optimized for detecting cold, low-luminosity exoplanets.

\subsection{Non-redundant aperture masking}\label{sec:nrm}

\begin{figure}[!t]
	\begin{center}
		\includegraphics[width=14.5 cm]{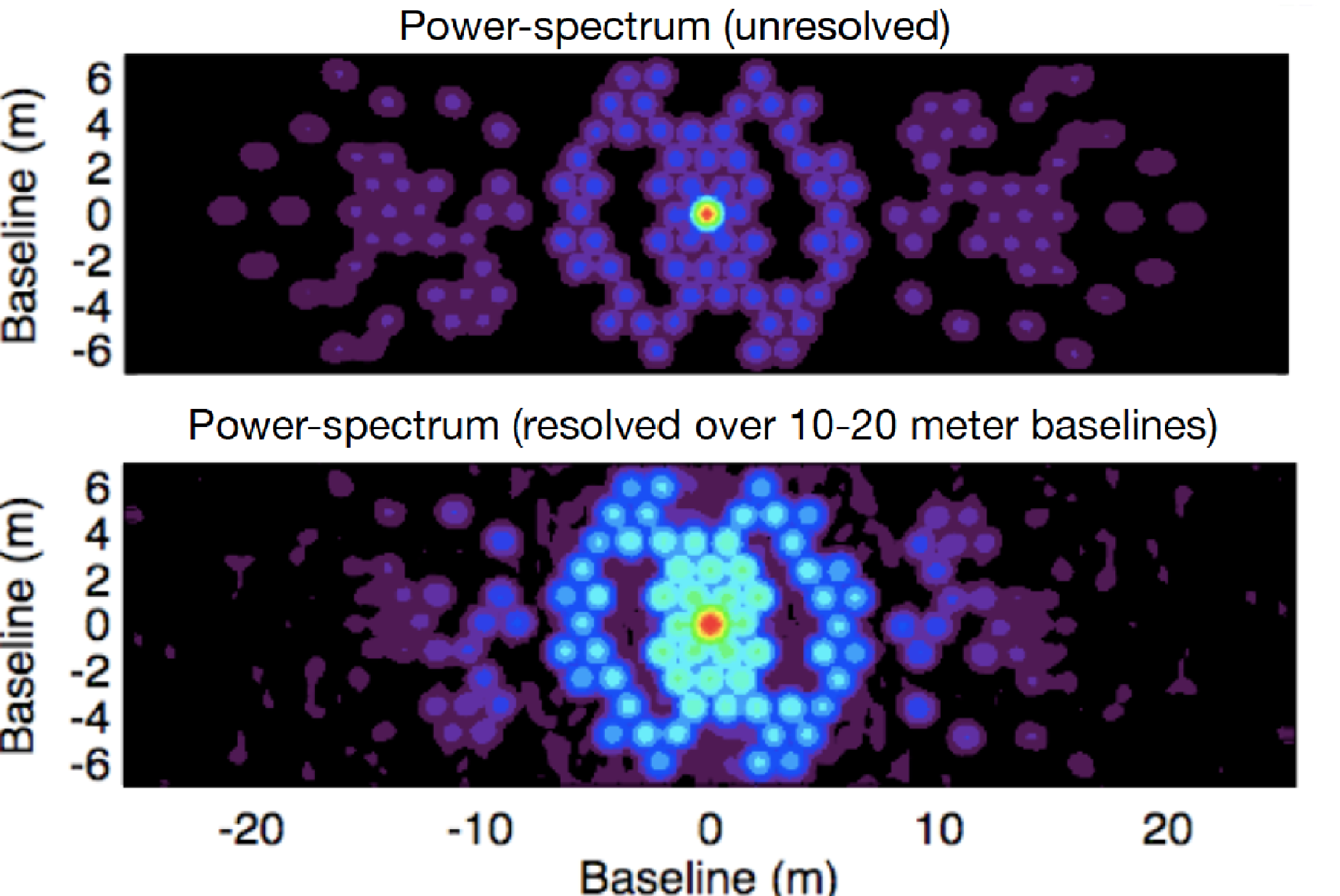}
		\caption{Example of non-redundant aperture masking observation with the LBTI. The figure shows the power spectrum of an unresolved calibrator (top) and a bright, previously imaged, YSO (bottom). The latter is heavily resolved especially on baselines longer than approximately 10m.} \label{fig:nrm}
	\end{center}
\end{figure}

The LMIRCam channel of the LBTI also contains two non-redundant aperture masks, one with 12 holes and one with 24 holes. An example of observation is shown in Figure~\ref{fig:nrm} for two relatively bright objects (L$<$4): an unresolved calibrator (top) and a bright, previously imaged, YSO (bottom) which is heavily resolved especially on baselines longer than about 10m. The observations were achieved without AO correction nor fringe tracking and a short enough integration time was used to freeze the atmospheric piston. AO correction and fringe-tracking are now available and will enable the same sort of data for fainter objects. Image reconstruction is currently under progress.


\subsection{Integral field spectrograph (ALES)}\label{sec:ales}

Integral field spectrographs are an important technology for exoplanet imaging, due to their ability to take spectra in a high-contrast environment, and improve planet detection sensitivity through spectral differential imaging\cite{Vigan:2010}. ALES (Arizona lenslet for exoplanet spectroscopy) is the first integral field spectrograph capable of imaging exoplanets from 3 to 5\,$\mu$m, and will extend our ability to characterize self-luminous exoplanets into a range where they peak in brightness. ALES was recently installed inside LBTI/LMIRcam as part of existing filter wheels, and first-light observations were obtained in June 2015. More information about ALES design and first-light observations are given in a dedicated paper in these proceedings\cite{Skemer:2015}.

\section{Summary and future work}\label{sec:summary}

The LBTI has achieved several instrumental milestones over the past few months including first wide-field Fizeau images, phase tracking, deep nulling interferometric observations, and first IFS images. Future work will be focused on improving the vibration environment and mitigating the impact of precipitable water vapor.  

\acknowledgments     
LBTI is funded by a NASA grant in support of the Exoplanet Exploration Program. LMIRCam is funded by the National Science Foundation through grant NSF AST-0705296. The LBT is an international collaboration among institutions in the United States, Italy and Germany. LBT Corporation partners are: The University of Arizona on behalf of the Arizona university system; Istituto Nazionale di Astrofisica, Italy; LBT Beteiligungsgesellschaft, Germany, representing the Max-Planck Society, the Astrophysical Institute Potsdam, and Heidelberg University; The Ohio State University, and The Research Corporation, on behalf of The University of Notre Dame, University of Minnesota and University of Virginia. 

\bibliographystyle{spiebib}   

\end{document}